\documentclass[aps,prd,a4paper,twocolumn,amsmath,showpacs,superscriptaddress,nofootinbib,preprintnumbers]{revtex4-1}

\usepackage{verbatim}
\usepackage[T1]{fontenc}
\usepackage[utf8]{inputenc}
\usepackage[american]{babel}
\usepackage{epsfig}
\usepackage{graphicx}
\usepackage{booktabs}
\usepackage{multirow}
\usepackage{dcolumn}
\usepackage{amsmath}
\usepackage{mathtools}
\usepackage{amsfonts}
\usepackage{amssymb}
\usepackage{epstopdf}
\usepackage{bm}
\usepackage{siunitx}
\usepackage{braket}
\usepackage{enumitem}
\usepackage{soul}
\usepackage[table]{xcolor}
\usepackage{color}
\usepackage{transparent}
\usepackage{pifont}
\definecolor{navyblue}{rgb}{0.0, 0.0, 0.5}
\definecolor{royalblue}{rgb}{0.25, 0.41, 0.88}
\definecolor{cadmiumgreen}{rgb}{0.0, 0.42, 0.24}
\definecolor{blue-violet}{rgb}{0.54, 0.17, 0.89}
\definecolor{darkviolet}{rgb}{0.58, 0.0, 0.83}
\definecolor{orange(colorwheel)}{rgb}{1.0, 0.5, 0.0}

\usepackage{hyperref}
\hypersetup{
    colorlinks=true, 
    linkcolor=royalblue, 
    citecolor=magenta}

\newcommand\ee{\end{equation}}
\newcommand\be{\begin{equation}}
\newcommand\eea{\end{eqnarray}}
\newcommand\bea{\begin{eqnarray}}

\newcommand{\msolar}{\textup{M}_\odot}

\usepackage{booktabs}
\usepackage{multirow}
\usepackage{dcolumn}
\usepackage{colortbl}

\definecolor{magenta(process)}{rgb}{1.0, 0.0, 0.56}

\definecolor{darkspringgreen}{rgb}{0.09, 0.45, 0.27}

\definecolor{royalblue(web)}{rgb}{0.25, 0.41, 0.88}

\begin{document}

\title{Do the early galaxies observed by JWST disagree with Planck's CMB polarization measurements?}

\author{Matteo Forconi}
\email{matteo.forconi@roma1.infn.it}
\affiliation{Physics Department and INFN, Universit\`a di Roma ``La Sapienza'', Ple Aldo Moro 2, 00185, Rome, Italy} 

\author{Ruchika}
\email{ruchika.ruchika@roma1.infn.it}
\affiliation{Physics Department and INFN, Universit\`a di Roma ``La Sapienza'', Ple Aldo Moro 2, 00185, Rome, Italy}

\author{Alessandro Melchiorri}
\email{alessandro.melchiorri@roma1.infn.it}
\affiliation{Physics Department and INFN, Universit\`a di Roma ``La Sapienza'', Ple Aldo Moro 2, 00185, Rome, Italy} 

\author{Olga Mena}
\email{omena@ific.uv.es}
\affiliation{Instituto de F{\'\i}sica Corpuscular  (CSIC-Universitat de Val{\`e}ncia), E-46980 Paterna, Spain}

\author{Nicola Menci}
\email{nicola.menci@oa-roma.inaf.it}
\affiliation{INAF - Osservatorio Astronomico di Roma, via Frascati 33, I-00078 Monte Porzio, Italy}

\date{\today}

\preprint{}
\begin{abstract}
The recent observations from the James Webb Space Telescope have led to a surprising discovery of a significant density of massive galaxies with masses of $M \ge 10^{10.5} M_{\odot}$ at redshifts of approximately $z\sim 10$. This corresponds to a stellar mass density of roughly $\rho_*\sim 10^6 M_{\odot} Mpc^{-3}$. Despite making conservative assumptions regarding galaxy formation, this finding may not be compatible with the standard $\Lambda$CDM cosmology that is favored by observations of CMB Anisotropies from the Planck satellite. In this paper, we confirm the substantial discrepancy with Planck's results within the $\Lambda$CDM framework. Assuming a value of $\epsilon=0.2$ for the efficiency of converting baryons into stars, we indeed find that the $\Lambda$CDM model is excluded at more than $99.7 \%$ confidence level (C.L.). An even more significant exclusion is found  for $\epsilon \sim 0.1$, while a better agreement, but still in tension at more than $95 \%$, is obtained for $\epsilon =0.32$.
This tension, as already discussed in the literature, could arise either from systematics in the JWST measurements or from new physics. Here, as a last-ditch effort, we point out that disregarding the large angular scale polarization obtained by Planck, which allows for significantly larger values of the matter clustering parameter $\sigma_8$, could lead to better agreement between Planck and JWST within the $\Lambda$CDM framework. Assuming $\Lambda$CDM and no systematics in the current JWST results, this implies either an unknown systematic error in current large angular scale CMB polarization measurements or an unidentified physical mechanism that could lower the expected amount of CMB polarization produced during the epoch of reionization. Interestingly, the model compatible with Planck temperature-only data and JWST observation also favors a higher Hubble constant $H_0=69.0\pm1.1$ km/s/Mpc at $68\%$ C.L., in better agreement with observations based on SN-Ia luminosity distances.
\end{abstract}

\maketitle

\section{Introduction}

Cosmology has seen remarkable progress in recent decades, driven by groundbreaking discoveries. These include cosmic acceleration identified through Type Ia Supernovae \cite{SupernovaSearchTeam:1998fmf,SupernovaCosmologyProject:1998vns}, precise measurements of Cosmic Microwave Background (CMB) temperature and polarization anisotropies \cite{Planck:2018vyg}, and the detection of Baryon Acoustic Oscillations through large-scale galaxy surveys \cite{SDSS:2005xqv}. At the core of these achievements is the $\Lambda$-Cold Dark Matter ($\Lambda$CDM) model, the standard cosmological model with six fundamental parameters, which has successfully explained various physical phenomena across different scales, including the aforementioned measurements.

However, the $\Lambda$CDM model is facing challenges from new observations, which has led to puzzles such as the Hubble constant tension~\cite{Riess:2022mme}. The Hubble constant tension refers to the inconsistency between the value of the Hubble constant obtained from Planck-2018 CMB observations ($H_0=67.4\pm0.5$ km~s$^{-1}$~Mpc$^{-1}$ from \cite{Planck:2018vyg}) and the value obtained from local measurements using Sn-Ia calibrated with Cepheid variable stars ($H_0=73\pm1$ km~s$^{-1}$~Mpc$^{-1}$ from ~\cite{Riess:2021jrx}). This persistent tension, that is now present above the 5 sigma level, has sparked debates about the validity of the $\Lambda$CDM model (see~\cite{DiValentino:2021izs,DiValentino:2020zio,Perivolaropoulos:2021jda,Abdalla:2022yfr,Verde:2019ivm,Riess:2019qba,DiValentino:2020vnx,DiValentino:2022fjm}).

Moreover, a second tension, named the $S_8$ tension, is now present
(see e.g. ~\cite{DiValentino:2020vvd}). The $S_8$ parameter is a combination of two important cosmological parameters, the amplitude of matter density fluctuations ($\sigma_8$) and the density of matter ($\Omega_m$), which together determine the overall clustering of matter in the Universe, such that $S_8=\sigma_8\sqrt{\Omega_m}$. $S_8$ could be inferred from the measurements of CMB anisotropies as those from Planck, and, more directly, from the measurements of galaxy lensing made by large surveys such as the Dark Energy Survey (DES) \cite{DES:2021vln,DES:2022ygi}, and the KiDS-1000 survey \cite{Heymans:2020gsg,KiDS:2020ghu}. Planck data favors a higher value of $S_8$, while galaxy surveys favour a lower value: these two measurements are not consistent with each other, showing a discrepancy at the 2-3 $\sigma$ statistical level.

It is therefore extremely important to further test the $\Lambda$CDM paradigm at different epochs and scales. One of the main challenges in testing the $\Lambda$CDM model was the lack of observations for objects at high redshift ($z\sim 10$). Current mainstream probes such as BAO and SNe Ia are unable to provide direct observations at these high redshifts, making it challenging to study the accuracy of the $\Lambda$CDM model around these epochs. This is a significant issue since it is during the early stages of the universe's evolution that the main structures of the universe are formed.

Recently, the James Webb Space Telescope (JWST) has observed galaxies at high redshifts, offering a glimpse into these 'dark ages'. These observations have revealed a population of surprisingly massive galaxy candidates (see e.g.~\cite{Santini_GLASS_Mstar,Castellano_GLASS_hiz,Finkelstein_CEERS_I1,Naidu_UVLF_2022,Treu:2022iti,Harikane_UVLFs,PerezGonzalez_CEERS, Papovich_CEERS_IV}) with stellar masses of the order of $M \ge 10^{10.5} M_{\odot}$. In a recent study \cite{2023Natu} it has been pointed out that the JWST data indicates a higher cumulative stellar mass density (CSMD) in the redshift range $7 < z < 11$ than predicted by the $\Lambda$CDM model, leading to several speculations that question the validity of the former~\cite{Boylan-Kolchin:2022kae,Biagetti:2022ode}. 

Nevertheless, the origin of the discrepancy could rely on other factors. One possibility to consider is that there may be inaccuracies in measuring the properties of galaxies. In fact, the stellar properties of the considered samples of massive high-redshift galaxies are derived from fitting a template of spectral energy distributions (SEDs) to the emissions in different photometric bands. The present lack of complete spectroscopic data raises the possibility of ambiguity in distinguishing between early star-forming galaxies and quiescent galaxies at lower redshifts, around $z \sim 5$. Additionally, the measured stellar masses strongly depend on the assumed Initial Mass Function (IMF). However, recent comparisons of photometric and spectroscopic redshifts in overlapping samples of galaxies with both measurements solidify the evidence for a high space density of bright galaxies at $z\gtrsim 8$ compared to theoretical model predictions \cite{2023arXiv230405378A}.

More severe uncertainties affect the measurements of stellar mass, which are derived assuming a Salpeter IMF. While adopting other universal forms for the IMF based on low-redshift conditions would not change (or may even amplify) the masses derived from the current measurements, the star formation process can be significantly different at high redshifts, resulting in a top-heavy IMF. In particular, the increase in gas temperatures in star-forming, high-redshift galaxies (also contributed by the heating due to CMB photons) could lead to a greater contribution of massive stars to the galactic light, resulting in significantly lower values for the stellar masses (a factor of 3-10, with the exact value depending on the assumed gas temperature) compared to those measured by Labbe' et al. (2022) \cite{2022arXiv220807879S}. However, recent studies have, for the first time, exploited the unique depth and resolution properties of JWST to perform spatially resolved SED fitting of galaxies in the SMACS 0723 JWST ERO Field, using six NIRCam imaging bands spanning the wavelength range 0.8–5 $\mu$m \cite{Gimenez-Arteaga:2022ubw}. This approach unveiled the presence of older stellar populations that are only distinguishable in the spatially resolved maps, since in the observations with single aperture photometry they are overshadowed by the younger ($10 Myr$)  population. As a result, the more accurate estimates of stellar masses derived from the resolved analyses are significantly larger (up to a factor of 10) than those obtained from the single-aperture photometry previously employed in the analysis of high-redshift galaxies observed with JWST. This would further strengthen the tension with LCDM predictions.

A second potential explanation is that the limited JWST observations thus far (covering an area of approximately $38$ square arcminutes) may be a highly atypical and unusually dense region of the Universe. This hypothesis can be tested through upcoming JWST surveys like COSMOS-Web.

If neither of the aforementioned possibilities can account for the discrepancies between the JWST results and the $\Lambda$CDM model, it may be necessary to consider modifications to the model itself.
Alternatives models within the dark matter and dark energy sector have been tested (for example, see ~\cite{Menci:2022wia,Wang:2022jvx,Gong:2022qjx,Zhitnitsky:2023znn,Dayal:2023nwi,Maio:2022lzg,Domenech:2023afs}), as well as Primordial Black Holes solutions~\cite{Liu:2022bvr,Yuan:2023bvh,Hutsi:2022fzw}, Cosmic strings~\cite{Jiao:2023wcn}, and large scale-dependence Non-Gaussianity~\cite{Biagetti:2022ode}. But also solutions within the $\Lambda$CDM paradigm have been studied~\cite{Haslbauer:2022vnq}, such as modifications of the primordial power spectrum~\cite{Parashari:2023cui} or using Extreme Value Statistics~\cite{Lovell:2}. Unfortunately, none of the extensions considered until present are capable of providing a compelling solution to the JWST stellar mass density observations.

While it is clearly timely and important to investigate the viability of extensions to the $\Lambda$CDM model, here we investigate another possible explanation, not explored yet in the literature, i.e. the presence of possible, unknown, systematics in the Planck data. The Planck data is widely regarded as one of the most reliable datasets in cosmology today. However, the measurement of large-scale polarization, in particular, has proven to be extremely challenging, as evidenced by the various constraints on the optical depth reported in different data releases (see e.g. \cite{Planck:2016mks}). The optical depth $\tau$ is directly linked to the amplitude of CMB polarization on large angular scales. For instance, the value of this parameter ranged from $\tau=0.089^{+0.012}_{-0.014}$ in the 2013 data release \cite{Planck:2013pxb} to $\tau=0.054^{+0.081}_{-0.071}$ in the 2018 data release (both at the $68\%$ confidence level). This $2.5$-sigma shift clearly highlights the difficulties in accurately determining the large-scale CMB polarization. Furthermore, the $TE$ spectrum at multipoles $ l<30$ has not been utilized for the extraction of cosmological parameters due to the existence of systematic uncertainties. The $TE$ spectra indeed exhibit excess variance compared to simulations at low multipoles, particularly at $\ell = 5$ and $\ell = 18$ and $\ell = 19$, for reasons that are not yet understood. At the very same time, Planck polarization measurements at high angular scales have also been plagued by systematics, such as temperature-to-polarization leakage and uncertainties in polarization efficiencies. As clearly stated in Ref.~\cite{Planck:2018vyg}, one should avoid over-interpreting the Planck polarization results, considering the sensitivity of those to small changes in specific choices and assumptions made in the data analyses. 

It is also worth noting that several physical effects exist that could alter the amplitude and shape of CMB polarization spectra, such as magnetic fields, interactions with pseudoscalar fields (Chern-Simons coupling), and axion-like particles (see e.g. \cite{Harari:1992ea, Pogosian:2019jbt,Lue:1998mq,Fujita:2020aqt,Finelli:2008jv}). Moreover, the modelling of the reionization process can in principle also affect the constraints from Planck on $\tau$ (see e.g. \cite{Mortonson:2007tb}), even if this is claimed to be small in the analysis of ~\cite{Planck:2018vyg}.

Taking a more conservative approach, we demonstrate that by excluding the constraints from CMB polarization data, a higher value of the $\sigma_8$ parameter becomes more compatible with the Planck data, leading to better agreement with the JWST observations. It should be noted that this solution is not entirely satisfactory as it exacerbates the tension with cosmic shear surveys, which favor a lower value of the $S_8$ parameter. Nevertheless, we show that this approach significantly alleviates the Hubble tension instead, which is currently one of the most challenging issues in cosmology.

The manuscript is organized as follows: in the next section, we describe our analysis method, including the derivation of the Cumulative Mass Stellar Density  given a cosmological model using the Press-Schechter formalism, and the extended models we considered. In Section \ref{sec:dataset}, we present the data used in our analysis. In Section \ref{sec:res}, we show our results, and in Section \ref{sec:conc}, we discuss our conclusions.

\section{Method}
\label{sec:method}

Our main observable is the Cumulative Stellar Mass Density (CSMD) given by

\begin{equation}
    \rho_\star(\bar{M})=\epsilon f_b\int^{z_2}_{z_1}\int^{\infty}_{\bar{M}}\frac{dn}{dM}MdM\frac{dV}{V(z_1,z_2)}~,
    \label{CSMD}
\end{equation}

\noindent where $f_b=\Omega_b/\Omega_m$ is the cosmic baryon fraction, $\epsilon$ is the efficiency of converting baryons into stars, $V$ is the comoving volume of the universe between redshift $z_2$ and $z_1$, given by $4/3\pi [R_2^3-R_1^3]$, $R_2$ 
and $R_1$ being the comoving radius at respective redshifts.

The comoving number density of the collapsed objects between a mass range from $M$ to $M+\Delta M$ at a certain redshift is given by

\begin{equation}
    \frac{dn(M,z)}{dM}=F(\nu)\frac{\rho_m}{M^2}\left\lvert\frac{d\ln\sigma(M,z)}{d\ln M}\right\rvert~,
    \label{HMassFunction}
\end{equation}

\noindent where $\rho_m$ is the comoving matter density of the universe ($\rho_m = \Omega_m\rho_{crit}$) and $\rho_{crit}$ is critical density of universe, $\rho_{crit}= 2.8\times10^{11} h^2 \msolar Mpc^{-3}$.

In what follows, we make use of the Sheth-Tormen Halo mass function \cite{Sheth:1999mn,Sheth:1999su}:

\begin{equation}
    F(\nu)=\nu f(\nu)=A\sqrt{\frac{2a}{\pi}}\nu\left(1+\frac{1}{\bar{\nu}^p}\right)e^{-\frac{\bar{\nu}^2}{2}}~,
\end{equation}

\noindent with $A=0.3222$, $a=0.707$, $p=0.6$. $\bar{\nu}=\sqrt{a}\nu$ and $\nu=\frac{\delta_c}{\sigma(R)}$. The parameter $\delta_c$ is the value of linear density contrast at the time of the collapse of non-linear density,  $\delta_c=1.686$ and $\sigma(R,z)$ is the variance of linear density field smoothed at scale R = (3M/ 4$\pi \bar{\rho})^{1/3}$:

\begin{equation}
    \sigma(R,z)^2=\frac{1}{2\pi^2}\int^\infty_0k^2P(k,z)W^2(kR)dk~,
\label{Variance}
\end{equation}

\noindent where $k$ is the comoving wavenumber of the matter power spectrum $P(k)$ and we assume here a Top-Hat window function given by
\begin{equation}
    W(kR)=3\frac{(\sin{kR}-kR\cos{kR})}{(kR)^3}~.
\end{equation}

Though it has been confirmed by high-quality N-body simulations that the accuracy of Sheth-Tormen (ST) Mass function is  and several phenomenological fitting mass functions may work better than ST Mass function \cite{basilakos,bhattacharya}, we choose to work within the ST formalism, as \emph{i} it is theoretically motivated in terms of the collapse of halos \cite{Maggiore_2010,Achitouv_2012}; and \emph{ii}  it has been exhaustively tested by N-body simulations for different dark energy models taking different priors for $\Omega_m$ and $\Omega_{\Lambda}$ \cite{despali_2015}. Various studies confirm that the ST mass function works universally as a function of redshift ($z<10$) and cosmology with 20 percent expected error bounds \cite{reed_2006,despali_2015}. Nevertheless, analyses of future  JWST data may need to consider a more sophisticated fit. 

\section{Datasets}
\label{sec:dataset}

The main goal this paper is to asses in a quantitative way the level of compatibility of the CSMD measurements from  JWST and the Planck measurements under the assumption of the $\Lambda$CDM model. For the Planck data, we consider the following CMB temperature and polarization power spectra from the final Planck release~\cite{Planck:2018vyg,Planck:2019nip}:

    \begin{itemize}
    
        \item The full Planck Temperature and Polarization anisotropies power spectra {\bf{Planck TTTEEE+low$\ell$+low$E$}}. 
        
        \item As above but excluding the large angular scale polarization ({\bf{Planck TTTEEE+low$\ell$}})
        
        \item  The full Temperature anisotropies spectra ({\bf{Planck TT+low$\ell$}})

        \item As above but considering only the small scale anisotropy ({\bf{Planck TT}})
    \end{itemize}

\noindent For the case of JWST's CSMD, we consider separately two sets of datapoints in two redshift bins respectively as taken from ~\cite{2023Natu}.

\noindent For the redshift bin $7<z<8.5$ we consider:

\begin{itemize}
    \item $\log_{10}\rho^{\rm obs}_*(M_1)=5.90\pm0.35$ at 
 $log_{10}(M_1)=10.1$ 
 \item $\log_{10}\rho^{\rm obs}_\star(M_2)=5.70\pm0.65$ at $log_{10}(M_2)=10.8$ 
 \end{itemize}

\noindent while for the redshift bin $8.5<z<10$ we exploit the following data points:

\begin{itemize}
    \item $\log_{10}\rho^{\rm obs}_\star(M_1)=5.7\pm0.40$ at 
 $log_{10}(M_1)=9.7$ 
 \item $\log_{10}\rho^{\rm obs}_\star(M_2)=5.40\pm0.65$ at $log_{10}(M_2)=10.4$ 
 \end{itemize}
 
\noindent In this paper, we therefore assume a Log Normal distribution for $\rho^{\rm obs}$, suggested by the symmetry of the lower and upper error bars under this choice. We also opt for this approach given the susceptibility of current measurements to substantial systematic errors and the exploratory character of our paper, which seeks to propose potential solutions rather than assert them.

Notice that the observational data points are model dependent because the comoving volume for $\rho^{obs}$ has been computed assuming the best-fit $\Lambda$CDM model to {\bf Planck TTTEEE+low$E$+lensing} CMB measurements ($h=0.6732$, $\Omega_m=0.3158$, $n_s=0.96605$, $\sigma_8=0.8120$, see ~\cite{2023Natu}). Therefore, to use these datapoints to analyze cosmologies different from the minimal $\Lambda$CDM  picture we need to rescale them properly by means of the comoving volume $V_C$ of the models under consideration. Given a cosmological model defined by set of cosmological parameters $\bar{p}$, we need to rescale the previous datapoints as 

\begin{equation}
\rho^{obs}(M_i,\bar p)={{V_C^{Planck}}\over{V_C(\bar p)}}\rho^{obs}_*(M_i)~.
\end{equation}

A similar rescaling is also applied using the square of the luminosity distance, due to the usage of the flux in the derivations of both $\rho^{obs}$ and the masses $M_i$ and, as before, the underlying assumption of the best-fit Planck $\Lambda$CDM model.

For our analyses, we perform a Monte Carlo Markov Chain (MCMC)  using the publicly available package \texttt{Cobaya} ~\cite{Torrado:2020dgo} and computing the theorical predictions exploiting the latest version of the publicly available software \texttt{CAMB}~\cite{Lewis:1999bs,2012JCAP}. We explore the posteriors of our parameter space using the MCMC sampler developed for CosmoMC~\cite{Lewis:2002ah,Lewis:2013hha} and tailored for parameter spaces with a speed hierarchy which also implements the ”fast dragging” procedure~\cite{neal2005taking}. The convergence of the chains obtained with this procedure is tested using the Gelman-Rubin criterion~\cite{Gelman:1992zz} and we choose as a threshold for chain convergence $R-1\lesssim 0.02$. Once the chains have converged, we compute the CMSD for each model, i.e. for each of the parameter combination explored by the MCMC analysis. In particular, we set the cosmology by varying the cosmological parameters $\{\Omega_b,\Omega_c,\theta_s,\tau\}$ and the inflationary parameters $\{n_s,A_s\}$. Afterwards, we use \texttt{CAMB} to compute the primordial power spectrum and subsequently the variance in Eq.~(\ref{Variance}). Then, we take its derivative and estimate the Halo mass function, see  Eq.~(\ref{HMassFunction}). Eventually, by double integrating over mass and redshift, Eq.~(\ref{CSMD}), we finally obtain $\rho_\star(M)$.

For each model in the chain, we include the JWST's CSMD data using  the (simple) $\chi^2$ function:
\begin{equation}
    \chi^2_{\textrm{JWST}}(\bar{p}) =  \sum^2_{i=1} \left[ \frac{
    \log_{10}{\rho^{\textrm{th}}(M_i,\bar{p})}- \log_{10}{\rho^{\textrm{obs}}(M_i,\bar{p})}}{\sigma_i} \right]^2~,
\end{equation}

\noindent where $\rho^{th}$ is the CSMD presented in Eq.~(\ref{CSMD}) for the different models explored and for an assumed valued of $\epsilon$, and $\sigma_i^2$ is the variance for each $\log_{10}\rho^{obs}(M_i,\bar{p})$ value \footnote{It is worth noting that one could assume free variation in $\epsilon$ and utilize the combined Planck+JWST data to impose constraints on this parameter. However, due to the potential existence of significant systematics, we opt for a more conservative approach. We choose to analyze only a few cases for this parameter, with a greater emphasis on understanding its variation rather than seeking stringent constraints on it.}. We then obtain the updated constraints on the cosmological parameters by re-weighting the MCMC chains, i.e. performing an importance sample, using the package \texttt{getdist}.

\section{Results}
\label{sec:res}
\begin{center}                                                        
\begin{table*}[htb]                                                   
\begin{tabular}{|l|c|c|}                                              
\hline\hline                                                          
Dataset & $\Delta \chi^2$ & $\Delta \chi^2$ \\ 
& $7\le z \le 8.5$ & $8.5 \le z \le 10$\\ 
\hline
\hline
$\epsilon=0.1$&&\\
\hline
Planck TT +JWST& $20.14$& $25.71$\\
Planck TT+Low$\ell$+JWST&$21.61$&$28.11$\\
Planck TTTEEE+Low$\ell$+JWST&$26.30$&$33.03$\\
Planck TTTEEE+Low$\ell$+LowE+JWST&$42.20$&$52.71$\\
\hline
\hline
$\epsilon=0.2$&&\\
\hline
Planck TT +JWST& $5.69$& $7.49$\\
Planck TT+Low$\ell$+JWST&$5.49$&$7.65$\\
Planck TTTEEE+Low$\ell$+JWST&$6.42$&$10.30$\\
Planck TTTEEE+Low$\ell$+LowE+JWST&$11.87$&$17.76$\\
\hline
\hline
$\epsilon=0.32$&&\\
\hline
Planck TT +JWST& $2.11$& $2.53$\\
Planck TT+Low$\ell$+JWST&$2.34$&$3.06$\\
Planck TTTEEE+Low$\ell$+JWST&$2.17$&$2.88$\\
Planck TTTEEE+Low$\ell$+LowE+JWST&$3.68$&$6.19$\\
\hline
\end{tabular}
\caption{ $\Delta \chi^2$ between the best fit model in the corresponding Planck and Planck+JWST chains.}\label{Table-chi2}       
\end{table*}                                                          
\end{center}                                                          

Table \ref{Table-chi2} presents the main results from our analyses, which is the goodness of fit of the $\Lambda$CDM models preferred by the Planck experiment compared to those from a combined Planck+JWST analysis. We consider three possible values for the efficiency parameter $\epsilon$ (see Eq.~(\ref{CSMD})): $\epsilon=0.1$, in agreement with current local observations; $\epsilon=0.2$, in agreement with high-redshift simulations; and a larger value of $\epsilon=0.32$, which serves as a conservative upper limit, see ~\cite{Tacchella:2018qny}.

When $\epsilon=0.1$, there is a strong incompatibility between the Planck CMB angular spectra measurements (even temperature-only) and the CMSD derived from JWST observations. A $\Delta \chi^2 \sim 10.6$ for two degrees of freedom, corresponding to the inclusion of the two JWST data points, already exceeds the $99.5\%$ confidence level (C.L.), indicating a high level of incompatibility between the datasets. Assuming that both datasets and the theory are correct, this result strongly suggests a value of $\epsilon > 0.1$. This is not surprising, as simulations predict larger values for $\epsilon$ at higher redshifts~\cite{Tacchella:2018qny}.

On the other hand, when assuming a value of $\epsilon=0.2$, there is a much better compatibility with the Planck temperature data alone, with a $\Delta \chi ^2 \le 5.9$ below the $95\%$ C.L. exclusion threshold for two degrees of freedom. Although the comparison at higher redshifts leads to a poorer fit, it is still reasonable since $\Delta \chi ^2 \le 7.4$ corresponds to an exclusion below the $97.5\%$ C.L. In the case of the low-redshift bin data, the inclusion of polarization data at small angular scales slightly increases the exclusion to just over $95\%$ C.L. However, the addition of Planck's large angular scale polarization ($LowE$) significantly raises the $\Delta \chi^2$ and excludes any compatibility below the $99.5\%$ C.L.
In conclusion, when $\epsilon=0.2$, a reasonable compatibility is achieved between the Planck temperature-only dataset and the CMSD derived from JWST. If this result is confirmed in the future, it could hint towards a systematic issue with the measurements of large-scale polarization data from Planck.

If we consider the case $\epsilon =0.32$, we observe that both datasets are relatively well-matched, with only the full Planck dataset exhibiting a tension above $95\%$ C.L. However, given the potential presence of various systematics in both datasets, this tension should not be regarded as a serious concern. This result may come as a surprise, as an efficiency of approximately $\epsilon=0.32$ is not inherently impossible, and, considering the errors associated with the CMSD, this finding is reasonable. In conclusion, if the large-scale polarization measured by Planck is accurate and the $\Lambda$CDM model is valid, the Planck vs JWST controversy can be resolved by a significant increase in the efficiency parameter $\epsilon>0.32$~\footnote{In REf.~\cite{Boylan-Kolchin:2022kae} the effect of $\epsilon$ is also explored. However, a full MCMC analysis with different possible data combinations, as the one presented here, was missing in the literature.}.

\begin{figure*}[htp]
	\centering
	\includegraphics[width=0.47 \textwidth]{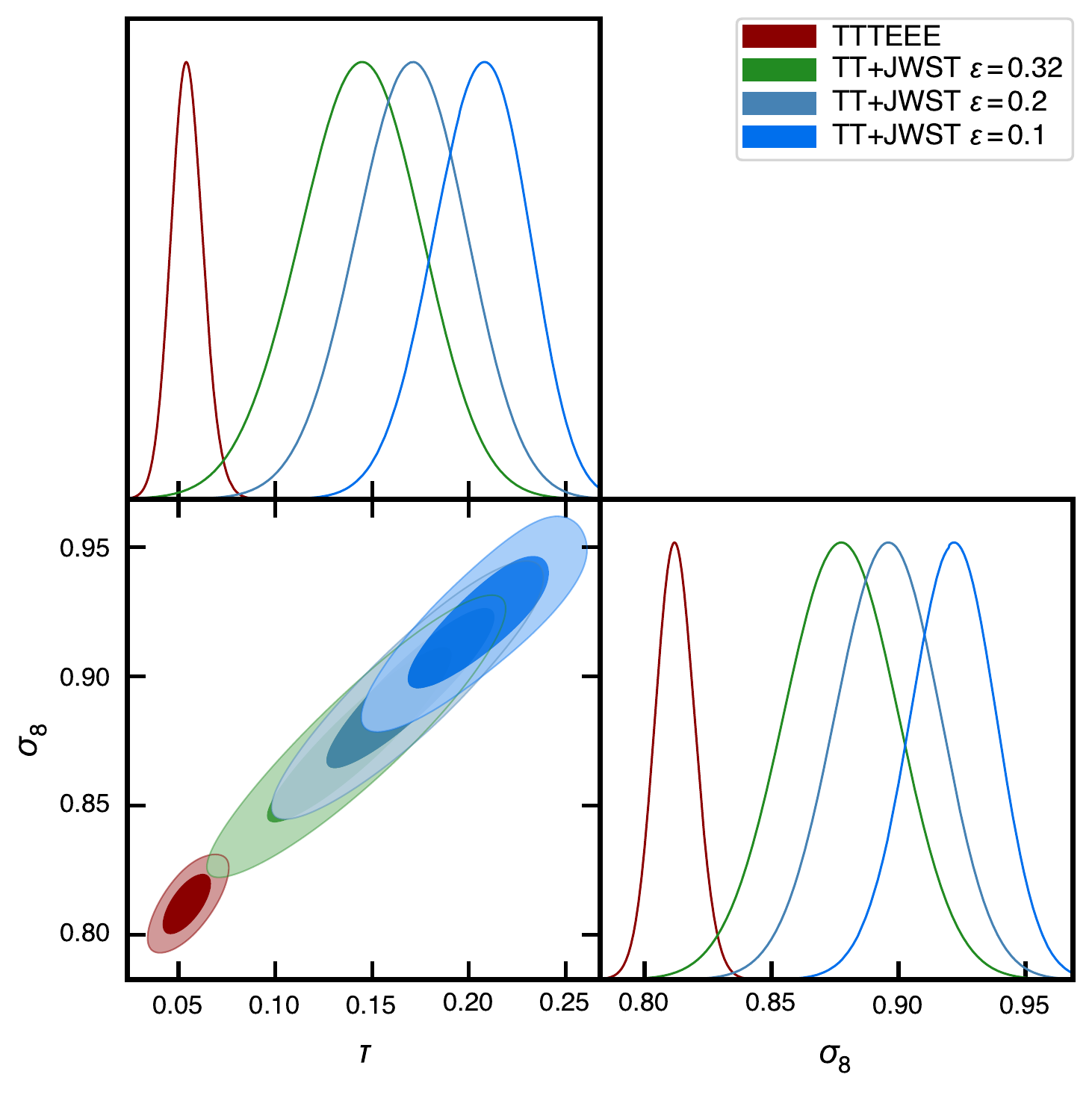}
 	\includegraphics[width=0.47 \textwidth]{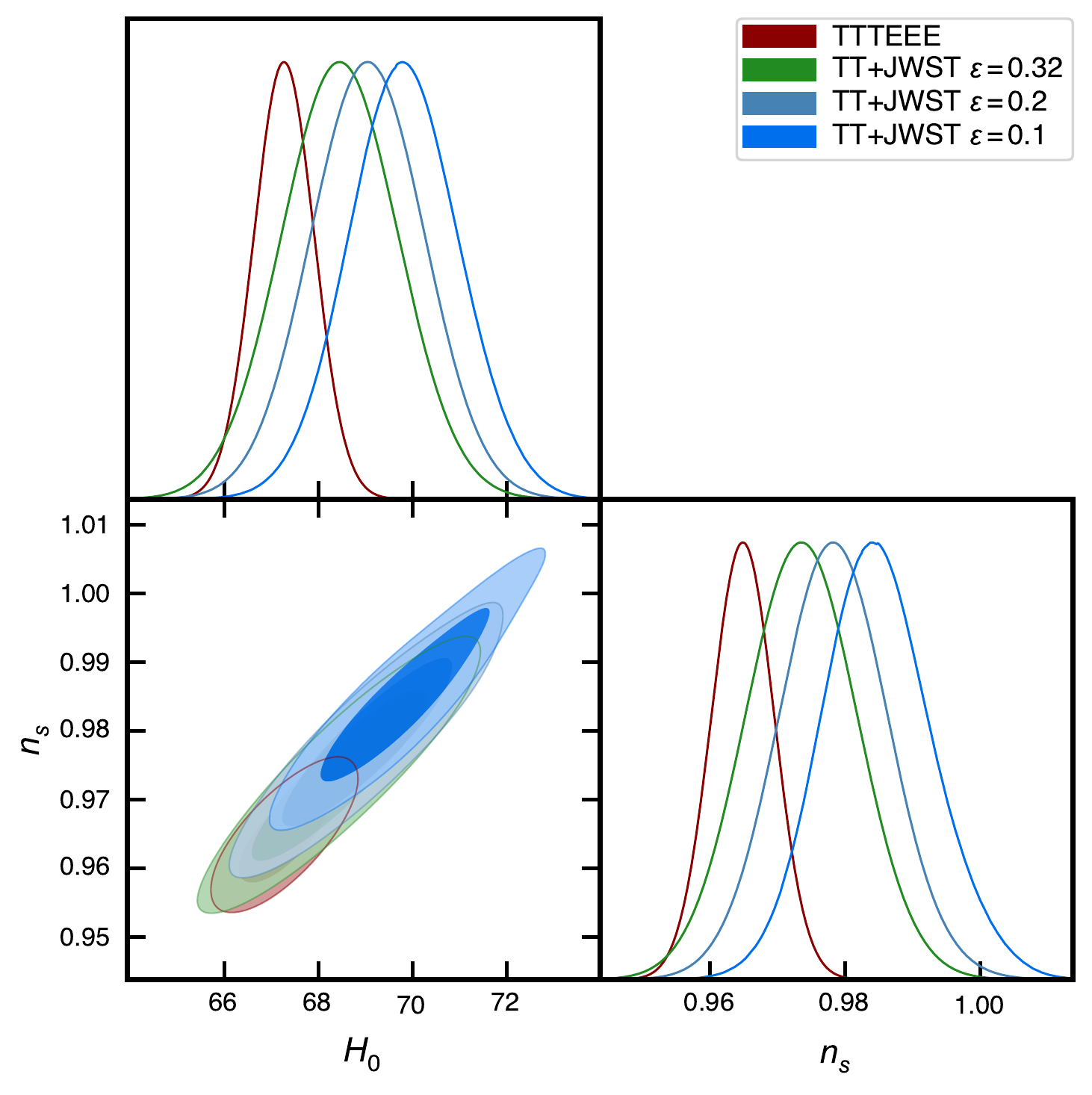}
	\caption{Marginalized 2D and 1D posterior distributions under the
 assumption of a $\Lambda$CDM model for the full Planck dataset and Planck TT+JWST in the range  $8.5\le z \le 10$. Constraints are reported in terms of the $\sigma_8$ and $\tau$ parameters (Left Panel) and in terms of the $n_S$ and $H_0$ parameters.} 
	\label{fig1}
\end{figure*}

Given the higher level of compatibility, it is interesting now to examine the constraints on cosmological parameters obtained from a combined analysis of Planck TT+JWST data and compare them with a standard analysis that utilizes the full Planck dataset (temperature and polarization).

Figure \ref{fig1} illustrates the comparison between a combined analysis of Planck TT+JWST data to a standard analysis that utilizes the full Planck dataset (temperature and polarization). In the left panel, we observe that the combined Planck TT+JWST analysis shifts the values of the $\sigma_8$ parameter towards higher values. This shift is made possible by the fact that the optical depth $\tau$ remains unconstrained without the inclusion of large-scale polarization data from Planck.

The 2D marginalized plot demonstrates that the full Planck dataset and the Planck TT+JWST dataset exhibit significant tension when $\epsilon \le 0.2$ and marginal compatibility when $\epsilon =0.32$. This tension can be quantified further by examining the constraints on the $\sigma_8$ parameter. For $\epsilon=0.2$, the Planck TT+JWST analysis yields a constraint of $\sigma_8=0.895\pm0.019$ at $68 \%$ C.L., which deviates from the $\sigma_8=0.8120\pm0.0073$ constraint derived from the full Planck dataset by $4.7$ standard deviations (see also the posteriors for $\sigma_8$ in the left panel of Fig.~\ref{fig1}). Such high value of the $\sigma_8$ parameter is certainly in tension with current determinations from cosmic shear data that actually prefer an even lower value ( see e.g. \cite{DES:2021vln,DES:2022ygi,Heymans:2020gsg,KiDS:2020ghu}). Lyman-$\alpha$ forest data, hovewer, shows generally an higher value for $\sigma_8\sim0.9$ (see e.g. \cite{2022MNRAS.515..857E}). 

Interestingly, as shown in the right panel of Fig.~\ref{fig1}, the Planck TT+JWST dataset also favors a higher value for the Hubble constant, with $H_0 = 68.8\pm 1.1$ in the range $7\le z \le8.5$ and
$H_0 = 69.0\pm 1.1$ in the range  $8.5\le z \le 10$ at $68 \%$ C.L. for $\epsilon=0.2$. The exclusion of the Planck large-scale CMB polarization not only potentially reconciles the Planck and JWST results but also brings them into agreement with current local measurements of the Hubble constant within a range of $3$ standard deviations. However, abandoning the lowE polarization does not completely resolve the tension, but rather reduces it to a level that is more consistent with statistical fluctuations. It is evident that if there is a systematic issue in the LowE polarization data, it can hinder the accurate identification of a theoretical solution to the Hubble tension.

It is also crucial to quantify the discrepancy in the optical depth $\tau$. Using the Planck TT+JWST data under the assumption of $\epsilon=0.2$ and considering the higher redshift bin, we obtained $\tau=0.17\pm0.027$, which is approximately $4$ standard deviations away from the full Planck result. This may seem highly significant, but a similar internal discrepancy between the TT and TTTEEE results is also observed in the case of $\Lambda$CDM, as seen in the more recent PR4 NPIPE analysis (\cite{Rosenberg:2022sdy}, see Table 1 in that paper). Therefore, a fluctuation of this magnitude is not entirely unexpected, considering the presence of similar internal discrepancies in the Planck data.

\section{Conclusions}
\label{sec:conc}
In this paper, we conducted a joint analysis of the latest observations of the CSMD from high redshift galaxies measured by the JWST satellite and the constraints on CMB anisotropies from the Planck experiment.

Once again, we would like to emphasize that due to the numerous assumptions involved and the potential for systematic uncertainties in the data, we cannot draw definitive conclusions at this stage. Therefore, our result should be considered valuable mainly for future analyses that will use less premature data.

We find that the efficiency of star production from baryons, represented by the parameter $\epsilon$, plays a crucial role in our analysis. Up to $\epsilon\sim0.3$, there is a strong disagreement between the full Planck CMB angular spectra measurements and the CSMD derived from JWST observations. This suggests that a value of $\epsilon > 0.3$ is necessary for compatibility, assuming the validity of the $\Lambda$CDM model and the absence of systematics in both datasets. Therefore, our analyses show that the tension between the full Planck dataset and the JWST dataset can potentially be eased by increasing the efficiency parameter $\epsilon$. When $\epsilon=0.32$, the tension becomes below $3$ standard deviations, offering a possible resolution to the discrepancy with higher efficiencies.

However, we found that if the Planck large angular scale EE data is omitted, there is a much better agreement between the Planck temperature data alone and the CSMD from JWST even for a lower efficiency of $\epsilon=0.2$ (as expected at such redshifts according to \cite{Tacchella:2018qny}. The combined analysis of Planck TT+JWST data indicates a shift towards higher values of the $\sigma_8$ parameter compared to the standard analysis using the full Planck dataset. This shift is possible thanks to the exclusion of large-scale polarization data. For $\epsilon=0.2$, the Planck TT+JWST analysis yields constraints on the $\sigma_8$ parameter that deviate from the constraints derived from the full Planck dataset by approximately $4$ standard deviations. Interestingly, the Planck TT+JWST analysis also provides a higher value for the Hubble constant of $H_0 = 69.0\pm 1.1$ at $68 \%$ C.L., suggesting a potential reconciliation with current local measurements. It is clearly interesting that both the JWST measurements and the luminosity distance measurements of \cite{Riess:2022mme} show a disagreement (under $\Lambda$CDM) with Planck polarization at low multipoles. 

It is also important to note that since the reionization of the universe is primarily caused by the energetic radiation emitted by the first generation of stars and galaxies, the higher the redshift of massive galaxies, the higher the expected redshift of reionization. Consequently, this would result in a larger polarization signal, which contrasts with Planck's observations. Considering the challenges associated with such measurements, one could argue that a possible (even partial) solution to the current tensions lies in a systematic issue with the Planck polarization data at low multipoles.

In this paper, we have restricted our comparison to the Planck CMB angular anisotropy and polarization power spectrum. However, it is evident that other datasets could be taken into consideration. Notably, combining the Planck temperature spectra with Planck lensing data would yield a value of $\tau$ consistent with the one derived from the full Planck temperature plus anisotropy spectra, albeit with larger error bars. Nevertheless, it's important to recognize that this constraint is also more dependent on the model and could significantly change when factors like variations in the dark energy equation of state $w$ or in the neutrino masses are considered.
Additionally, it's worth noting that other cosmological datasets, like those obtained from high-redshift Ly-$\alpha$ observations (see e.g. \cite{Esposito:2022plo} and references therein), suggest a higher value of $\sigma_8\sim0.9$.
We plan to investigate additional models and datasets in a forthcoming paper.

The future Litebird satellite \cite{LiteBIRD:2020khw,Sugai:2020pjw}, scheduled for launching around 2030, holds great promise for accurately measuring the large-scale CMB polarization. This will provide a crucial test of the Planck results and shed further light on the tension observed between the full Planck dataset and the Planck TT+JWST dataset. By obtaining precise measurements of the large-scale CMB polarization, Litebird has the potential to validate or challenge the exclusion of Planck's large-scale CMB polarization in the combined analysis.

Additionally, the upcoming Euclid satellite \cite{EUCLID:2011zbd}, anticipated to be launched in the coming weeks, is expected to provide valuable insights into the $S_8$ tension and the nature of dark matter clustering. If the Euclid mission confirms the $S_8$ tension observed between the full Planck dataset and the current cosmic shear data, it could also exclude the possibility of a dark matter clustering scenario where $\sigma_8 \sim 0.9$. In summary, if future results from JWST are confirmed, the combination of precise CMB polarization measurements from Litebird and cosmic shear measurements from Euclid will significantly contribute in resolving the current tension between Planck and Planck TT+JWST observations.

\acknowledgments 
We would like to thank Mike Boylan-Kolchin, Marco Castellano, Sujatha Ramakrishnan, and Paola Santini for their valuable advice and insightful discussions. MF, R and AM are supported by TASP, iniziativa specifica INFN. This work has been partially supported by the MCIN/AEI/10.13039/501100011033 of Spain under grant PID2020-113644GB-I00 and by the European Union’s Framework Programme for Research and Innovation Horizon 2020 (2014–2020) under grant H2020-MSCA-ITN-2019/860881-HIDDeN.



\bibliography{Bibliography} 
\end{document}